\documentclass[%
 aps,
 prl,
 showpacs,
 amsmath,amssymb,
 reprint,%
groupedaddress
]{revtex4-1}

\usepackage{graphicx}
\usepackage{dcolumn}
\usepackage{upgreek}
\usepackage{bm}

\begin{document}


\title{Efficient expulsion of magnetic flux in superconducting RF cavities for high $Q_0$ applications}

\author{S. Posen}
 \email{sposen@fnal.gov}
\author{M. Checchin}
\author{A. C. Crawford}
\author{A. Grassellino}
\author{M. Martinello}
\author{O. Melnychuk}
\author{A. Romanenko}
\author{D. Sergatskov}

\affiliation{Fermi National Accelerator Laboratory, Batavia, Illinois, 60510, USA.}

\date{\today}

\begin{abstract}
Even when cooled through its transition temperature in the presence of an external magnetic field, a superconductor can expel nearly all external magnetic flux. This Letter presents an experimental study to identify the parameters that most strongly influence flux trapping in high purity niobium during cooldown. This is critical to the operation of superconducting radiofrequency cavities, in which trapped flux degrades the quality factor and therefore cryogenic efficiency. Flux expulsion was measured on a large survey of 1.3 GHz cavities prepared in various ways. It is shown that both spatial thermal gradient and high temperature treatment are critical to expelling external magnetic fields, while surface treatment has minimal effect. For the first time, it is shown that a cavity can be converted from poor expulsion behavior to strong expulsion behavior after furnace treatment, resulting in a substantial improvement in quality factor. Future plans are described to build on this result in order to optimize treatment for future cavities.
\end{abstract}

\maketitle

Recently, concentrated research effort has been devoted to obtaining high quality factors ($Q_0$) in superconducting radiofrequency (SRF) cavities, structures that transfer energy to beams in particle accelerators. High $Q_0$ reduces the considerable costs for cryogenics---both infrastructure and AC wall power for the cryogenic plant---required to cool cavities operating with high duty factor. Treatments such as nitrogen-doping \cite{Grassellino2013b} have been invented to substantially improve nominal quality factors, but $Q_0$ can be strongly degraded by trapped magnetic flux.

$Q_0$ degradation by trapped flux can be considered as a three step process: 1) the cavity is cooled in a finite external magnetic field environment $B_{ext}$; 2) some of that field, $B_{trap}$, is trapped in the surface of the cavity; and 3) the surface resistance $R_s$ of the cavity ($Q_0$ is inversely proportional to $R_s$) is increased by an amount $R_{fl}$ due to the trapped field. Preparation can be optimized to reduce the impact of each of these steps. Use of non-magnetic components, magnetic shielding, and active field cancellation can reduce $B_{ext}$ in step 1 (see e.g. Ref. \onlinecite{Edwards1995}). The mean free path can be optimized to reduce $R_{fl}$ for a given $B_{trap}$ in step 3 (see e.g. Refs. \onlinecite{Martinello2015,Gonnella2015}). For step 2, the amount of external field trapped in the cavity during cooldown can be reduced, and recent experiments have shown that even full expulsion is possible (previous experimental results \cite{Padamsee2008,Vallet1992} had reported full trapping---$B_{trap} \sim B_{ext}$). These recent experiments include the discovery that the fraction of external field that is trapped in the surface of a niobium cavity during cooldown is strongly dependent on the thermal gradient over its surface \cite{Romanenko2014,Romanenko2014a}, which may be explained by thermal forces \cite{Huebener1969} or other mechanisms \cite{Kubo2016}. In this Letter, we present an experimental study that further develops the understanding of flux expulsion in niobium cavities. For the first time, expulsion is studied as a function of both thermal gradient and cavity preparation. The goal of the study was to determine whether flux trapping behavior is determined by bulk properties (e.g. grain boundaries, as in \cite{Dasgupta1978,Santhanam1976}) or surface properties (e.g. nitrides from the nitrogen-doping process) and then to find a treatment that improves expulsion.


\begin{figure}[htbp]
\begin{center}
\includegraphics[width=0.48\textwidth,angle=0]{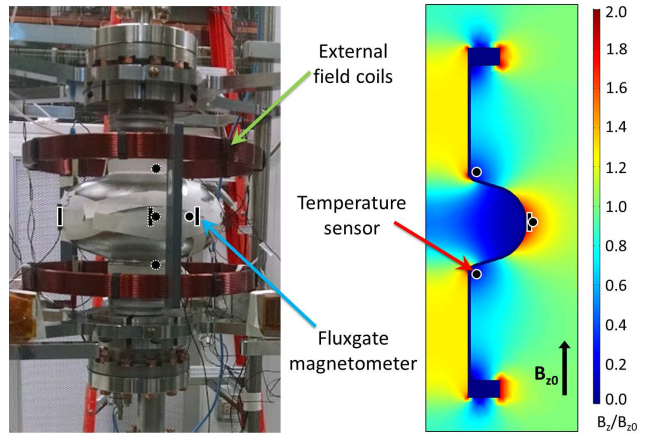}
\end{center}
\caption{Apparatus for measuring magnetic flux expulsion in a 1.3 GHz single cell SRF cavity (left), and simulation of axial field component resulting from the complete expulsion of an external field parallel to the cavity axis, normalized to the external field (right).}
\label{fig:apparatus}
\end{figure}

\begin{figure}[htbp]
\begin{center}
\includegraphics[width=0.36\textwidth,angle=0]{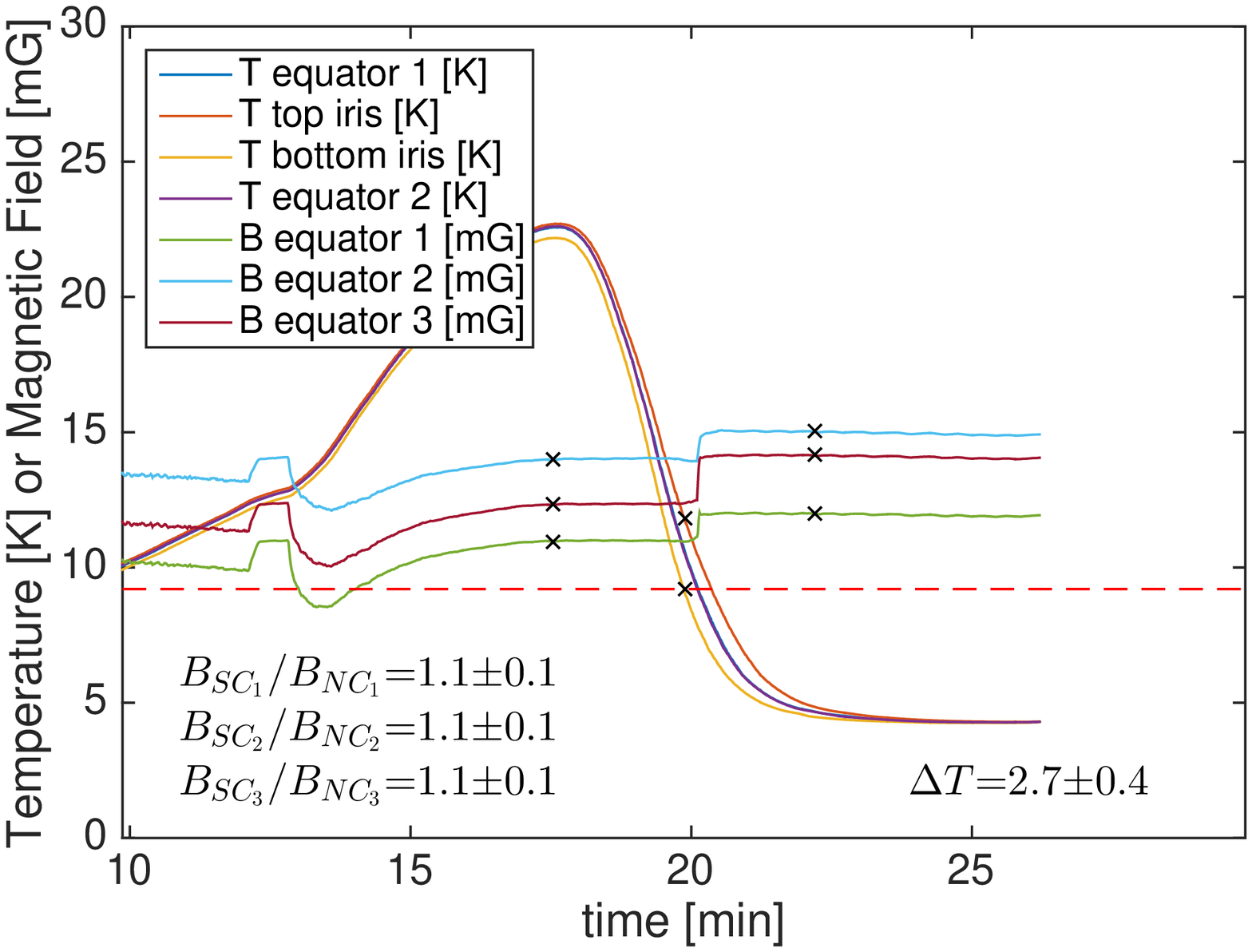}
\includegraphics[width=0.36\textwidth,angle=0]{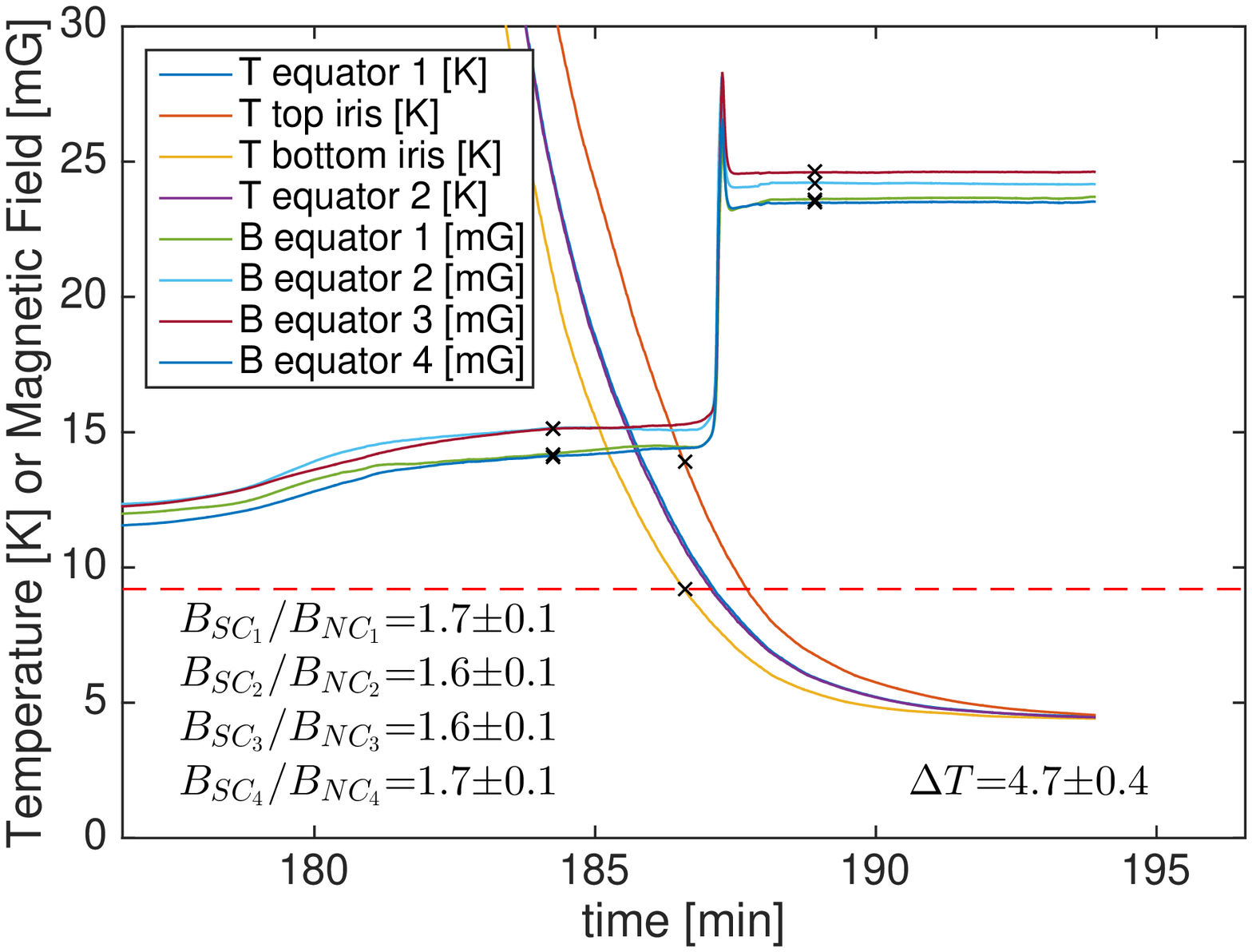}
\end{center}
\caption{Typical flux expulsion measurements. As the cavity passes through $T_c$ during cooldown (dashed line), temperature and magnetic field are recorded (illustrated with `x' symbols). Some cavity preparations can result in strong flux trapping behavior, showing $B_{SC}/B_{NC}$ ratios close to 1 for modest $\Delta T$ (e.g. top) and others result in efficient flux expulsion, with ratios close to 1.7 under similar conditions (e.g. bottom).}
\label{fig:expelvstime}
\end{figure}

The arrangement for measuring flux expulsion in a single cell cavity is shown in Figure \ref{fig:apparatus} (measurement technique from Ref. \onlinecite{Romanenko2014}). Magnetic field coils are arranged around the cavity, and a current is applied to create a field of ~10 mG at the surface. Fluxgate magnetometers are attached to the middle, oriented in the same direction as the applied field. Thermometers measure the temperature at the top and bottom of the cavity cell and the middle. During cooldown, as the temperature falls below the transition temperature $T_c$ and the cavity goes from the normal conducting (NC) state to the superconducting (SC) state, a step change is observed in the magnetic field sensors. The magnitude of this change corresponds to the amount of flux expelled. If $B_{ext}$ is fully trapped, the field measured above $T_c$, $B_{NC}$, is the same as the field measured below $T_c$, $B_{SC}$ ($B_{SC}/B_{NC}$=1). When $B_{ext}$ is completely expelled, calculations of the full Meissner expulsion show that the expected ratio of $B_{SC}/B_{NC}$ should be 1.7 \footnote{The calculation of $B_{SC}/B_{NC}$ for full expulsion takes into account the thickness of the fluxgate probe but not its length, and it assumes that the probe is ideally centered at the equator. Calculations show that the effect of these factors is expected to be $<5\%$, and they are omitted in the figures for simplicity.}. In this way, the measurement of $B_{SC}/B_{NC}$ reveals what fraction of flux is trapped during cooldown instead of being expelled. The temperature difference between the top and bottom thermometer at $T_c$ gives a measure of the spatial temperature gradient across the cavity. Typical measurements of flux expulsion are shown in Figure \ref{fig:expelvstime} \footnote{Sources of measurement uncertainty in $\Delta T$ include thermal impedance between cavity and thermometer, assumed to result in overall uncertainty of 0.4 K. Sources of measurement uncertainty in $B_{SC}/B_{NC}$ include radial background fields and misalignment of fluxgate probe, assumed to result in overall uncertainty of 0.1.}.


\begin{figure}[htbp]
\begin{center}
\includegraphics[width=0.36\textwidth,angle=0]{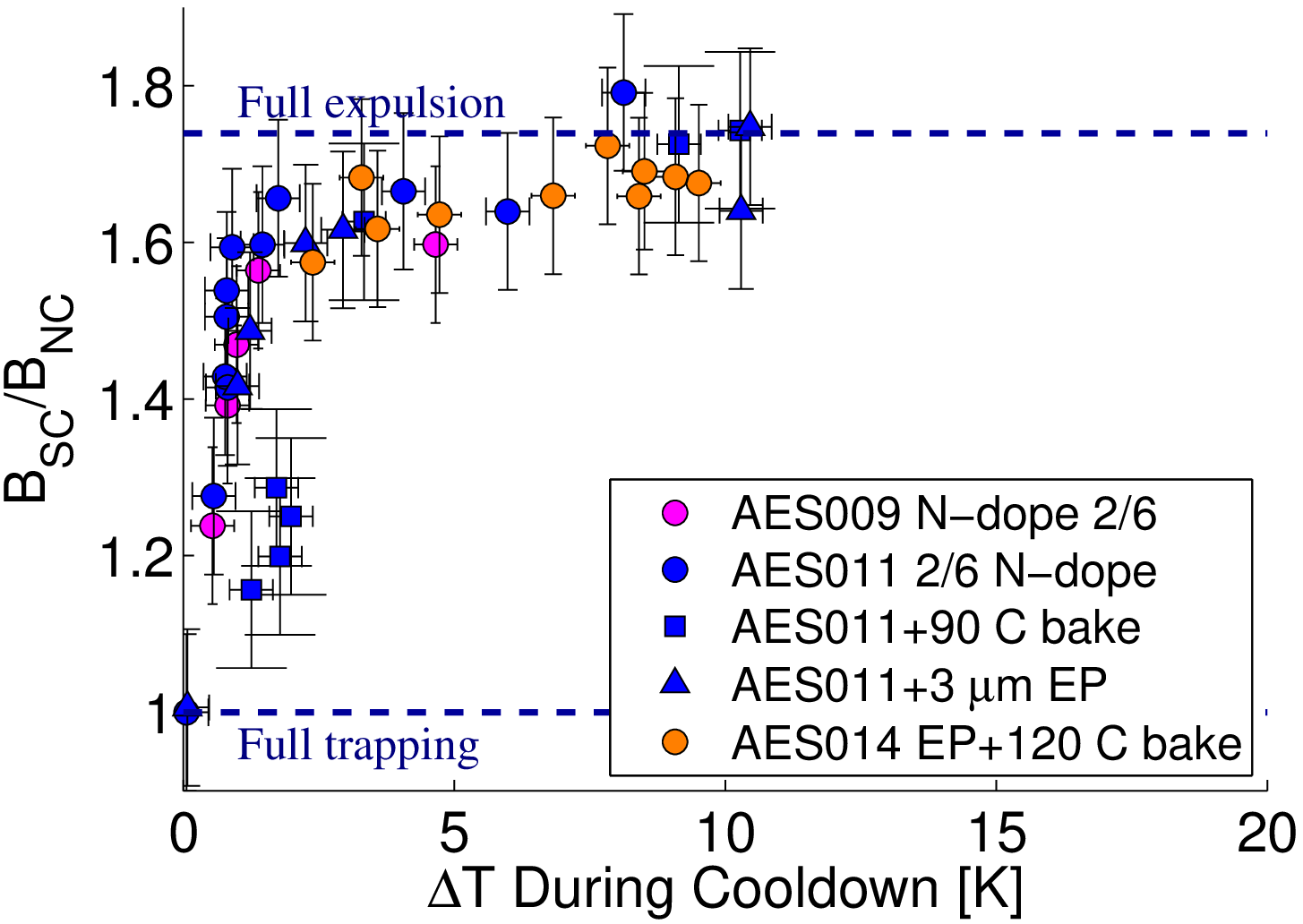}
\includegraphics[width=0.36\textwidth,angle=0]{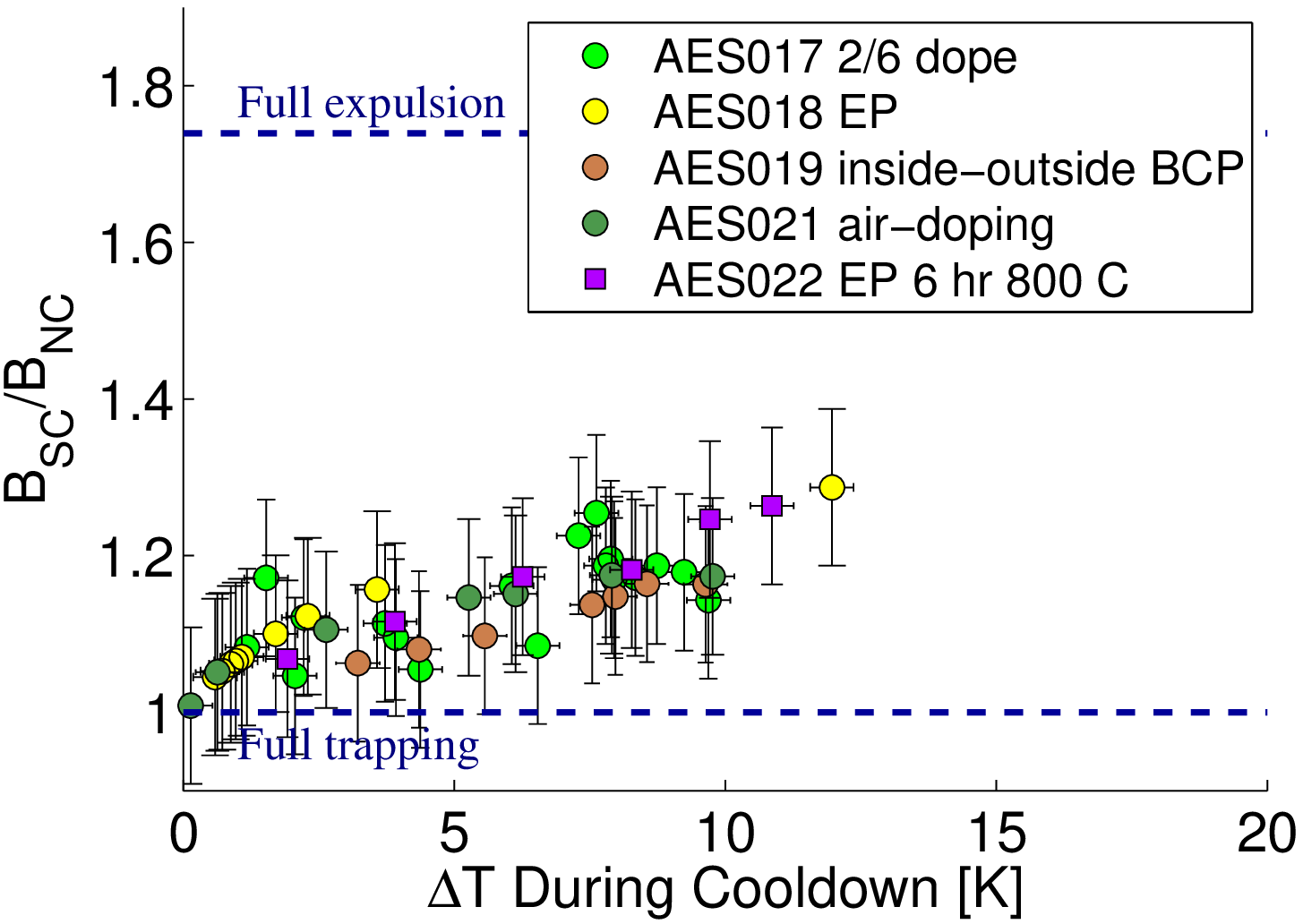}
\end{center}
\caption{Measured curves of flux expulsion as a function of temperature difference $\Delta T$ from bottom to top of the cavity cell as the cavity passes through $T_c$ during cooldown. The cavities measured from production group 1 (AES007-AES016) showed strong expulsion behavior (top) while those from production group 2 (AES017-AES022) showed strong trapping (bottom).}
\label{fig:batches}
\end{figure}

Several single cell 1.3 GHz cavities were measured over several cooldown cycles to show the trend with spatial temperature gradient for a given cavity preparation. A total of 22 datasets were measured, each for a different treatment. Each dataset consists of many cooldowns to ~7 K, varying $\Delta T$ \footnote{Higher starting temperatures generally led to larger $\Delta T$}. A trend in the data quickly became apparent, as shown in Figure \ref{fig:batches}. Two production groups of cavities from the same vendor had consistently different trapping behavior: the cavities from production group 1 expelled well while those in production group 2 expelled poorly. For the cavities that expel flux well, $\Delta T$ as low as ~2 K over the cavity cell are sufficient to expel the majority of the external field. For the cavities that expel poorly, the majority of the flux is trapped even for $\Delta T$ close to 10 K.

\begin{figure}[htbp]
\begin{center}
\includegraphics[width=0.36\textwidth,angle=0]{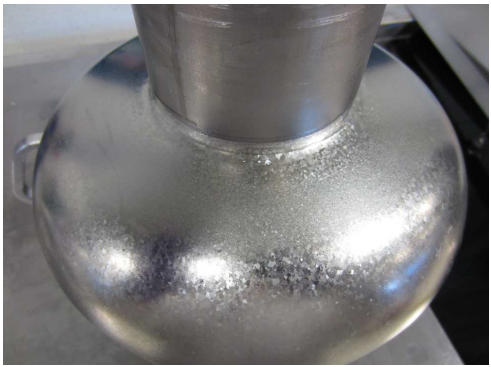}
\end{center}
\caption{Grain growth in AES011. The cavity was fabricated from material with $\sim$100 $\upmu$m sized grains, some of which grew to the few mm-scale after only a few UHV furnace cycles at 800$^\circ$C that were each 3 hours long or shorter.}
\label{fig:AES011grains}
\end{figure}

The cavities in production group 1 were acquired earlier and had previously undergone several rounds of processing, and they showed another notable behavior. After a few cycles of ultra-high-vacuum (UHV) furnace treatment at 800$^\circ$C for up to 3 hours, the cavities in this production group, which initially had grain size $\sim$100 $\upmu$m, showed strong grain growth. This is shown in Figure \ref{fig:AES011grains}.

Previous experiments on niobium samples studied the difference between fine grain and large grain material in fraction of flux trapped during cooldown \cite{Aull2012,Dhavale2012}. These studies were performed before it was recognized that it was important to control for thermal gradient, making it difficult to extract quantitative data, but qualitative trends were demonstrated. Material with larger grains appeared to have higher expulsion, suggesting that grain boundaries may act as pinning sites for flux, giving an advantage to cavities with fewer grain boundaries. This is consistent with the results in Figure \ref{fig:batches}, as well as with previous studies of high field pinning \cite{Dasgupta1978,Santhanam1976}. However, even in single crystal niobium samples, heat treatment appeared to improve expulsion, suggesting that e.g. dislocation content plays an important role \cite{Aull2012}.

\begin{figure}[htbp]
\begin{center}
\includegraphics[width=0.48\textwidth,angle=0]{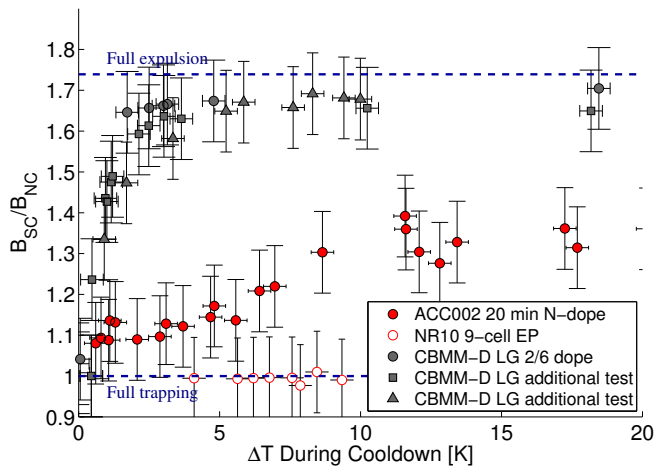}
\end{center}
\caption{Flux expulsion measurement in two 1.3 GHz fine grain cavities, single cell ACC002 and 9-cell NR010, and one large grain 1.3 GHz single cell, CBMM-D. It should be noted that CBMM-D received more furnace cycles than ACC002 or NR010.}
\label{fig:nonAES}
\end{figure}

\begin{figure}[htbp]
\begin{center}
\includegraphics[width=0.48\textwidth,angle=0]{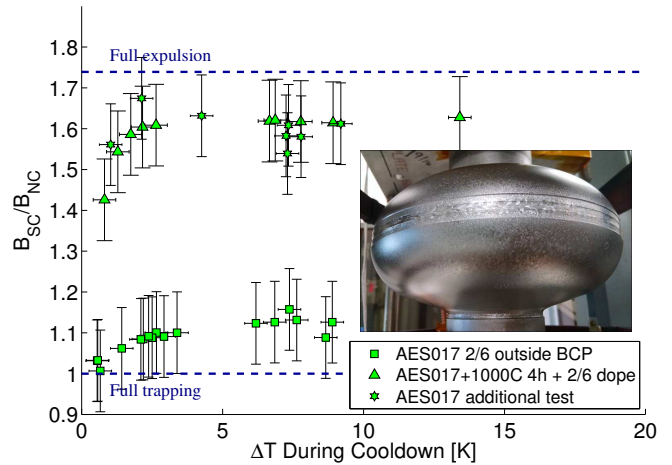}
\end{center}
\caption{AES017, a cavity from production group 2 that showed strong flux trapping behavior, was converted to expel strongly after a 1000$^\circ$C 4 h furnace treatment. The inset image shows the grain growth after treatment.}
\label{fig:AES017}
\end{figure}

Cavities from other production groups also show results consistent with this. Figure \ref{fig:nonAES} shows two fine grain cavities that expel poorly and one large grain cavity that expels well. To confirm the effect of bulk characteristics such as grain size and dislocation content in flux trapping, one of the cavities from production group 2 was heated at 1000$^\circ$C for 4 hours in a UHV furnace. The grain size was visibly increased and the expulsion improved substantially, as shown in Figure \ref{fig:AES017}. This strongly supports the hypothesis that bulk properties determine flux expulsion behavior.

\begin{figure}[htbp]
\begin{center}
\includegraphics[width=0.23\textwidth,angle=0]{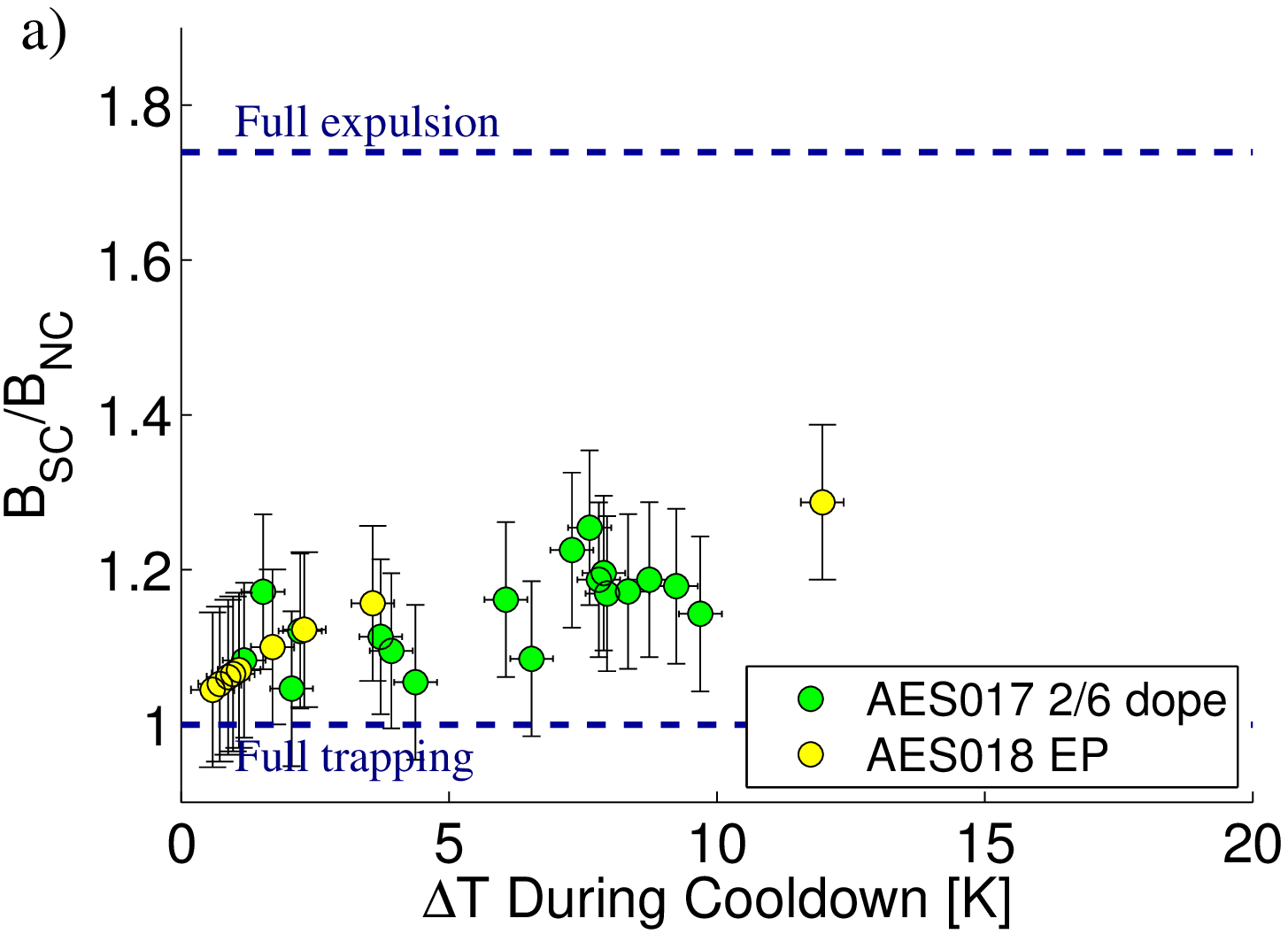}
\includegraphics[width=0.23\textwidth,angle=0]{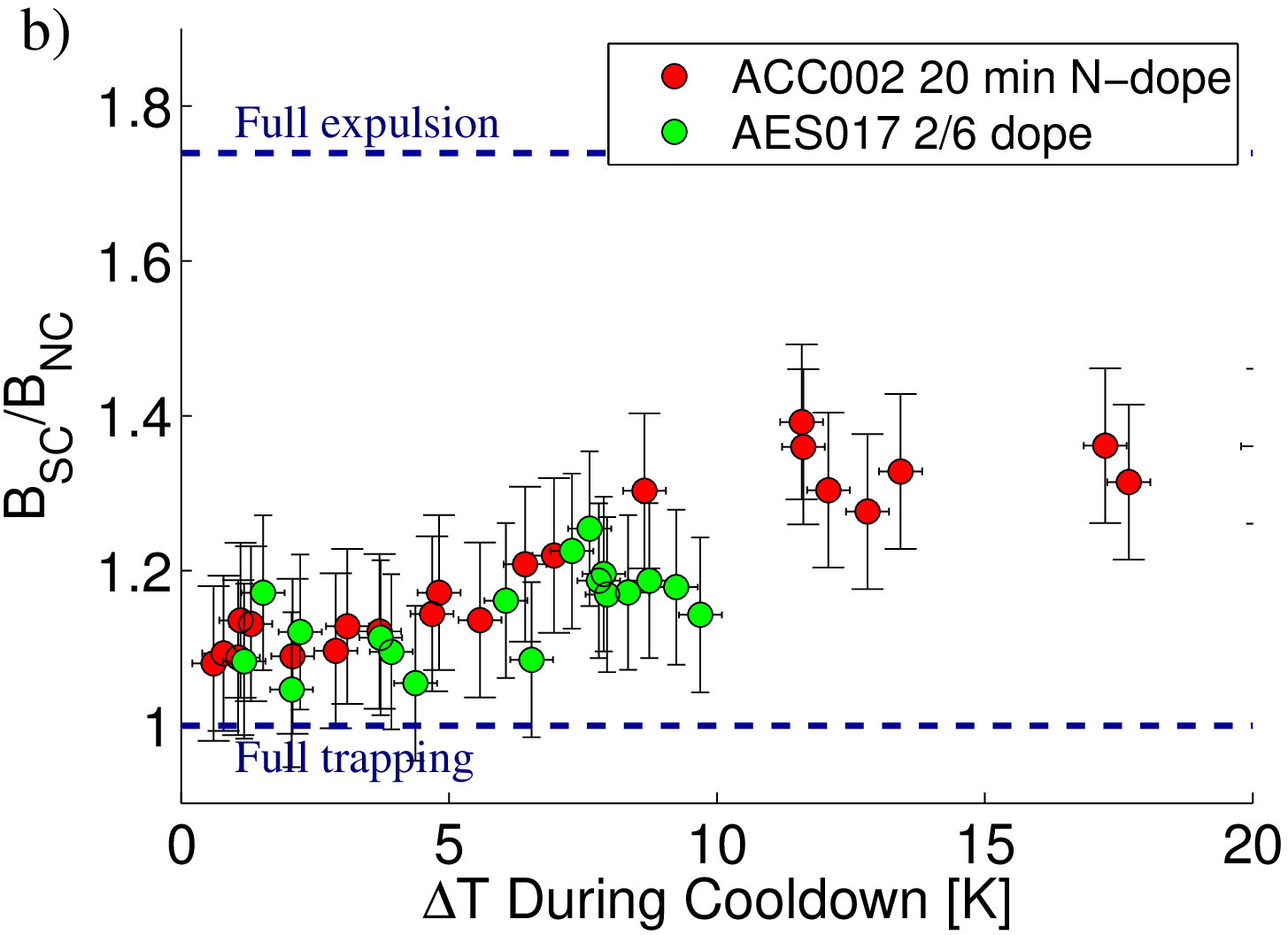}
\includegraphics[width=0.23\textwidth,angle=0]{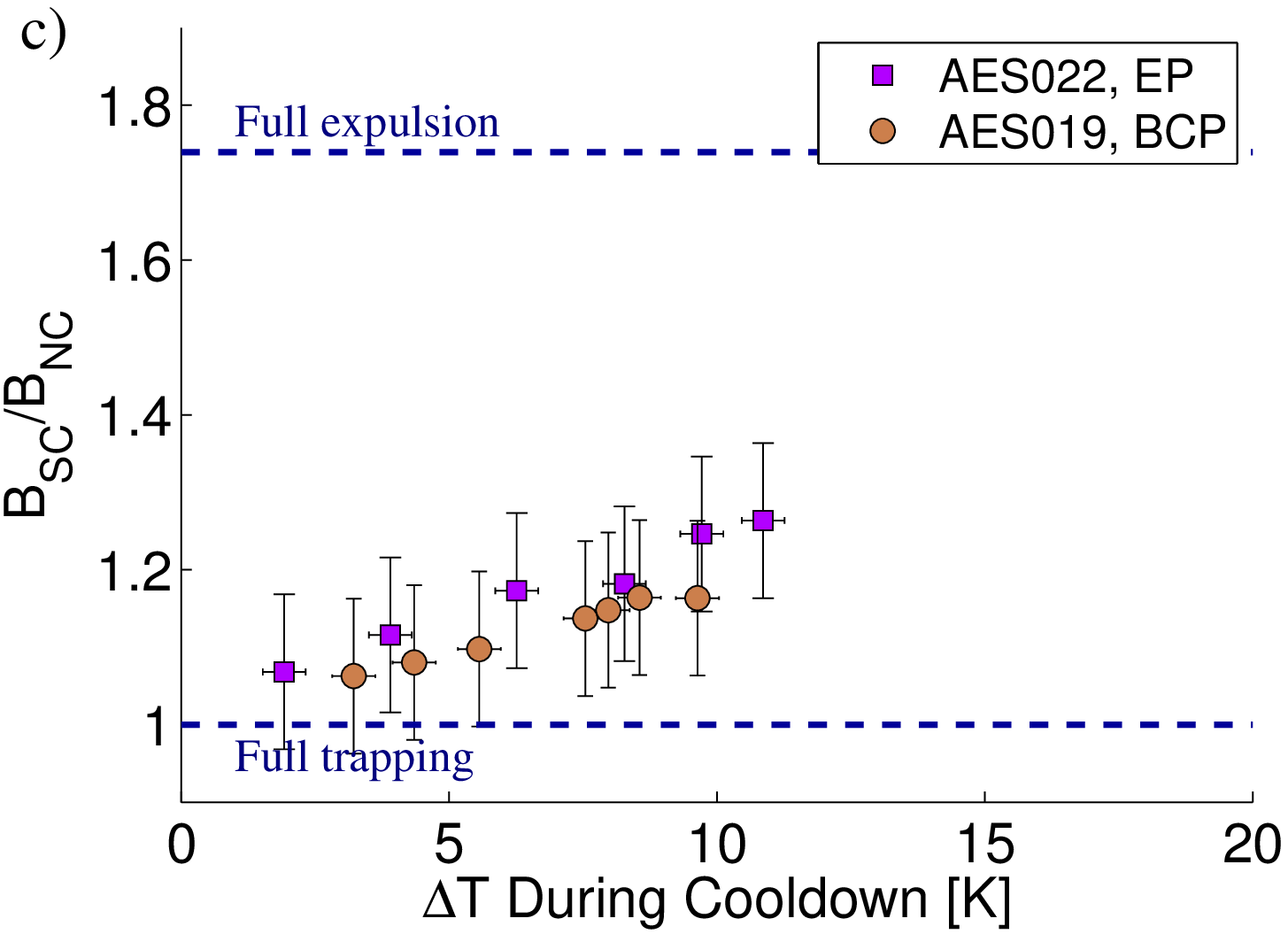}
\includegraphics[width=0.23\textwidth,angle=0]{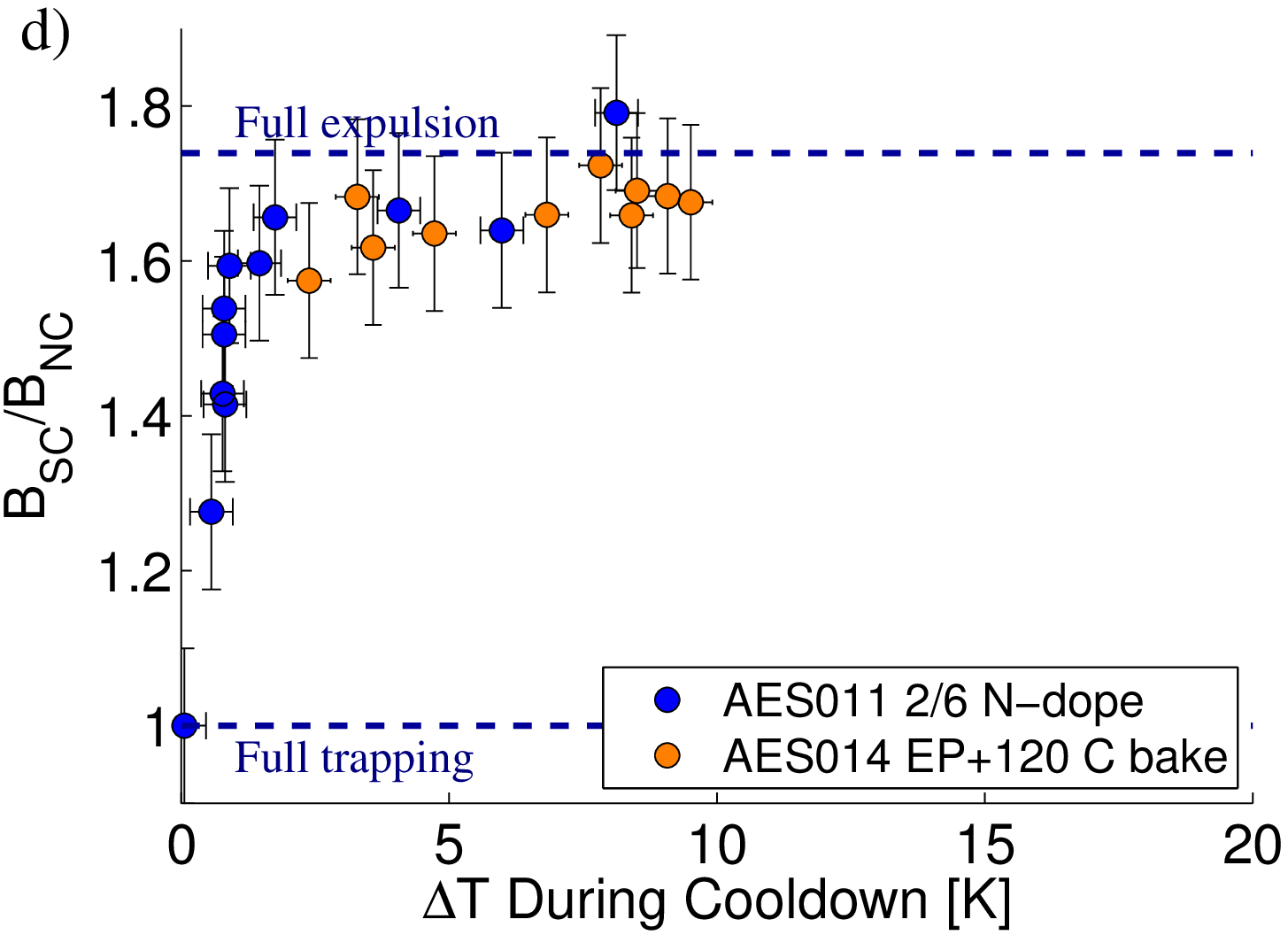}
\includegraphics[width=0.23\textwidth,angle=0]{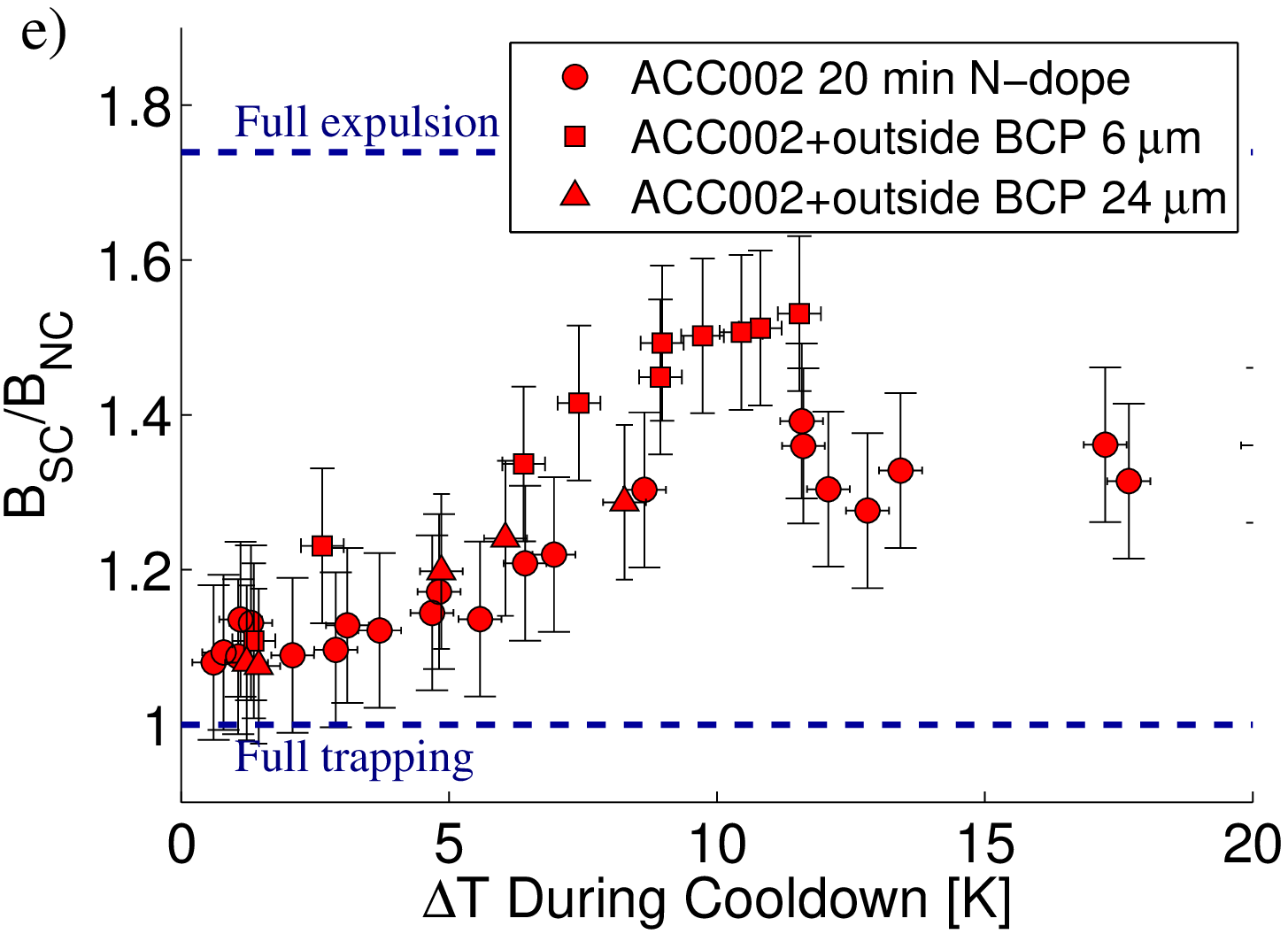}
\includegraphics[width=0.23\textwidth,angle=0]{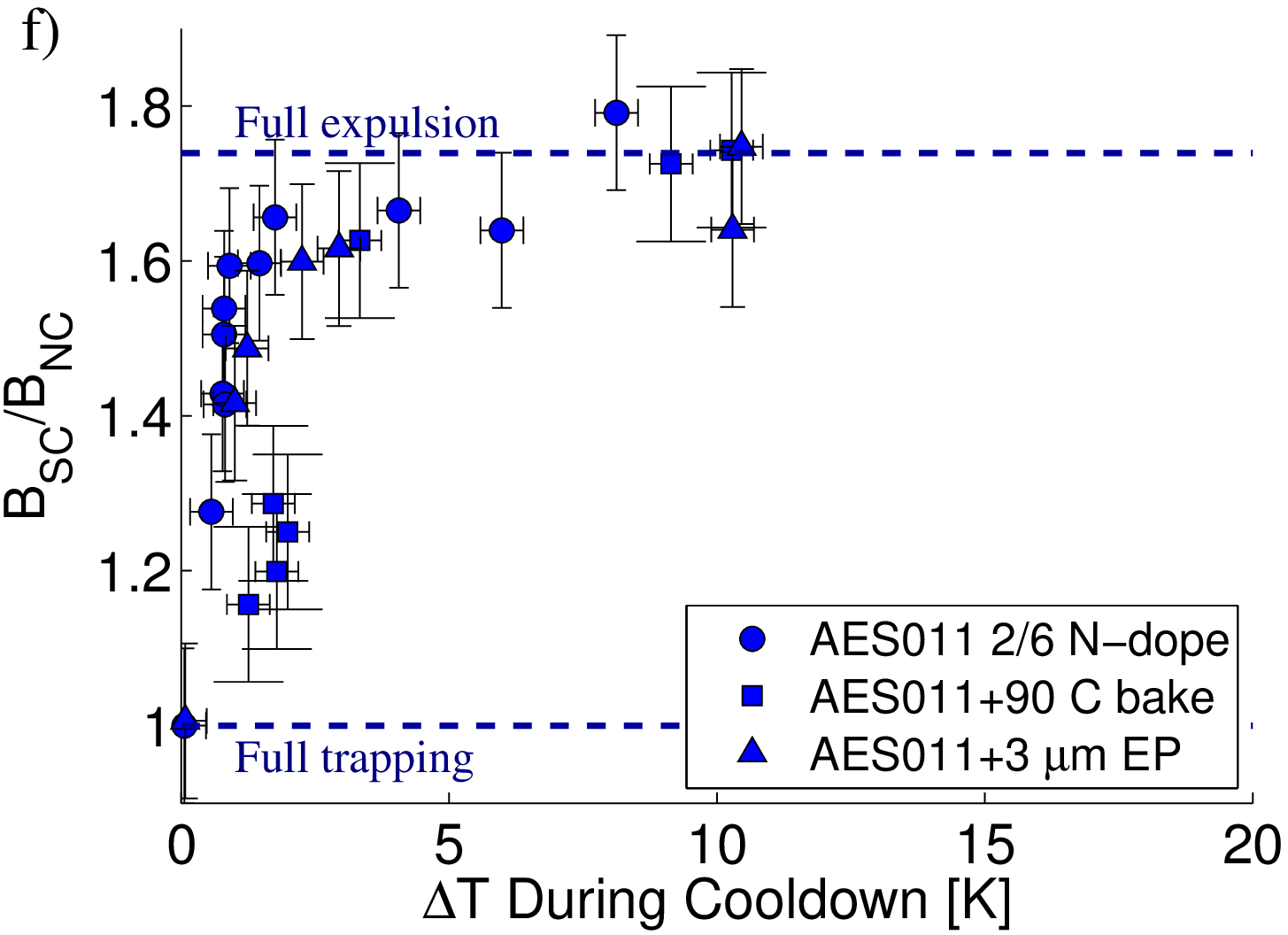}
\end{center}
\caption{Flux expulsion vs $\Delta T$, comparing a variety of different surface treatments with similar bulk history. Each graph compares curves for 2 or 3 treatments that produce very different surfaces but show similar expulsion behavior: a) EP vs N-doped; b) `light' N-dope (20 minutes in N at 800$^\circ$C) vs `heavy' N-dope (2 minutes in N at 800$^\circ$C + 6 minute anneal); c) EP vs BCP; d) light doping + external BCP vs EP + 120$^\circ$C bake; e) effect of light and heavy BCP of external surface f) effect of 90$^\circ$C bake and additional EP.}
\label{fig:sixplot}
\end{figure}

The cavities measured in the survey had been treated with a wide variety of surface processing techniques. By comparing cavities from the same production group, with similar furnace treatment history but different surface processing, we can study the effect of the surface on flux expulsion. We can also study the effect of a given surface treatment by comparing flux expulsion on a single cavity before and after treatment. Figure \ref{fig:sixplot} shows a number of such comparisons, such as electropolished (EP) surface vs buffered chemical polish (BCP) surface, N-doping with 2/6 recipe \cite{Grassellino2013b} vs EP, and as-treated outside surface vs outside BCP. In each case, the flux expulsion is nearly the same for cavities with similar bulk history regardless of surface conditions.

\begin{figure}[htbp]
\begin{center}
\includegraphics[width=0.36\textwidth,angle=0]{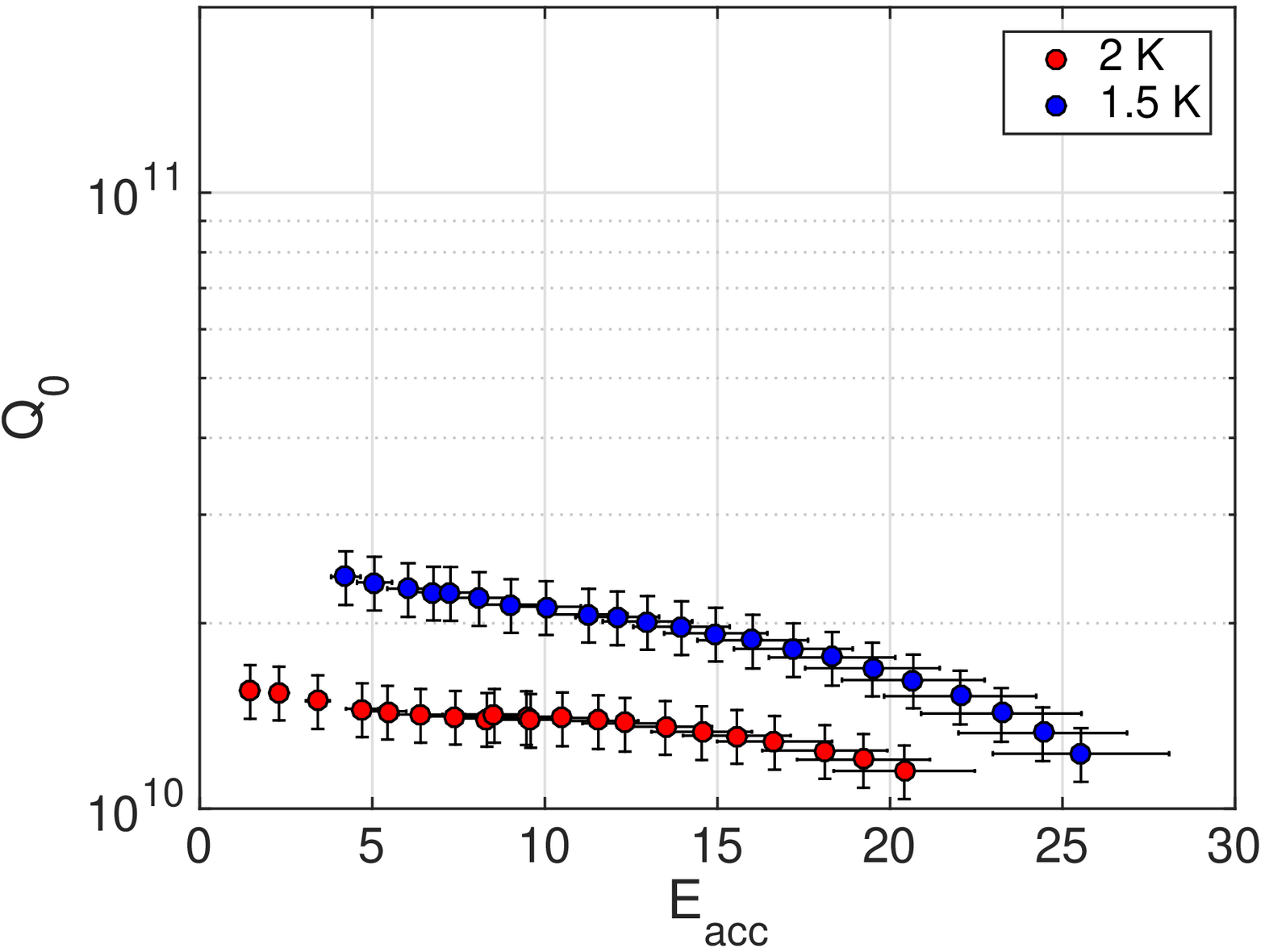}
\includegraphics[width=0.36\textwidth,angle=0]{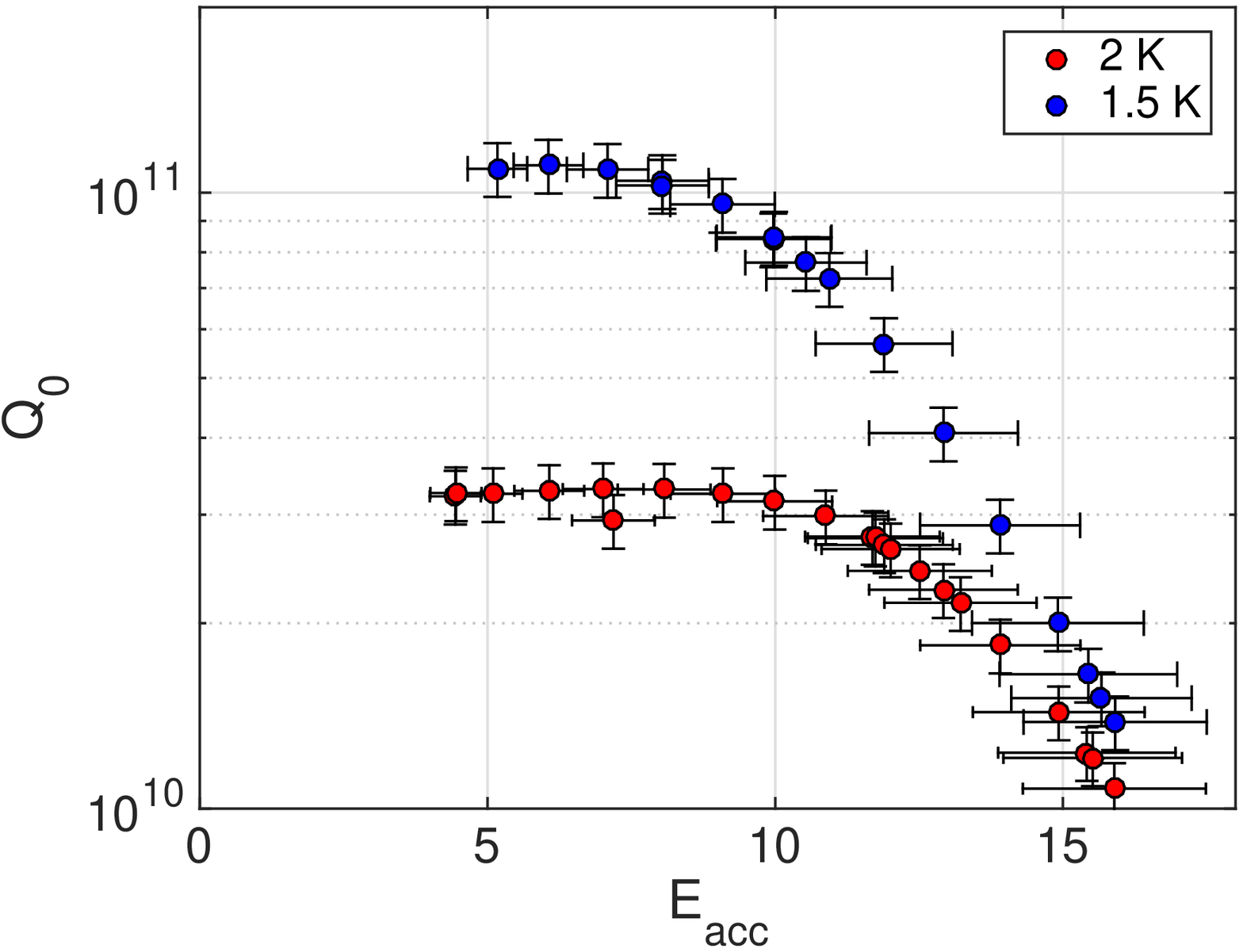}
\end{center}
\caption{$Q_0$ vs accelerating gradient $E_{acc}$ for AES017 after cooldown in a 10 mG field with a modest $\Delta T \sim$2-4 K. Before 1000$^\circ$C heat treatment (top), the cavity traps most of the flux and resulting in a low field $Q_0$ at 1.5 K of $\sim 2\times10^{10}$. After the heat treatment to improve flux expulsion (bottom), this value improves to $\sim 1\times10^{11}$.}
\label{fig:QvsE}
\end{figure}

To show that flux expulsion significantly affects $Q_0$, RF measurements were performed on the cavity from Figure \ref{fig:AES017} before and after 1000$^\circ$C furnace treatment. Each time, the cavity was cooled down in a 10 mG field with a modest $\Delta T$ of 2-4 K. The $B_{SC}/B_{NC}$ ratio measured before heat treatment was 1.1, showing that most of the external flux was trapped, while the ratio measured after was 1.6, showing strong expulsion (these cooldowns are shown in Figure \ref{fig:expelvstime}). Figure \ref{fig:QvsE} shows the corresponding substantial improvement in $Q_0$ at 1.5 K and 2 K \footnote{The substantial $Q$-slope observed after furnace treatment is not expected to be related to the flux expulsion mechanism (based on similar cavities that expelled well) nor to the grain growth (based on experience with large grain cavities). Rather, it is believed to be due to contamination from the furnace, based on observations of other cavities treated in a similar timeframe.}.

In this Letter, we have presented new results measuring expulsion of magnetic flux during cooldown through $T_c$ in niobium SRF cavities. It was found that efficient flux expulsion is strongly influenced not only by spatial thermal gradient but also by treatment. Cavities that showed signs of modified bulk structure after high temperature furnace treatment exhibited significantly stronger flux expulsion as a function of temperature than those that did not. The surface preparation showed no significant effect. Using these results, a procedure was designed that was shown to substantially improve flux expulsion behavior by UHV furnace treatment at 1000 C for 4 h. A 1.3 GHz cavity was evaluated before and after this procedure by cooling in a field $\sim$10 mG with a modest temperature gradient $\sim$2-4 K. After 1000 C treatment, RF measurements showed that $Q_0$ at 1.5 K improved by a factor of $\sim$5. Future studies will focus on determining what specific properties determine expulsion behavior---e.g. grain boundary density or dislocation content---and optimizing treatment for improving flux expulsion without compromising mechanical properties. For example, if grain boundary density is important to expulsion behavior, future cavities could be manufactured from material with larger grain size, or cavities could be given an optimized furnace treatment after manufacture to ensure that high $Q_0$ can be maintained in a cryomodule environment with a realistic background magnetic field. The experimental results presented here may be useful in other applications where magnetic field isolation is important, such as in quantum computing.

This work was supported by the United States Department of Energy, Offices of High Energy and Nuclear Physics. The authors are grateful for technical assistance from the FNAL cavity preparation and cryogenic teams. Fermilab is operated by Fermi Research Alliance, LLC under Contract No. DE-AC02-07CH11359 with the United States Department of Energy.

\bibliography{../../bibTeX/library}

\begin{thebibliography}{18}%
\makeatletter
\providecommand \@ifxundefined [1]{%
 \@ifx{#1\undefined}
}%
\providecommand \@ifnum [1]{%
 \ifnum #1\expandafter \@firstoftwo
 \else \expandafter \@secondoftwo
 \fi
}%
\providecommand \@ifx [1]{%
 \ifx #1\expandafter \@firstoftwo
 \else \expandafter \@secondoftwo
 \fi
}%
\providecommand \natexlab [1]{#1}%
\providecommand \enquote  [1]{``#1''}%
\providecommand \bibnamefont  [1]{#1}%
\providecommand \bibfnamefont [1]{#1}%
\providecommand \citenamefont [1]{#1}%
\providecommand \href@noop [0]{\@secondoftwo}%
\providecommand \href [0]{\begingroup \@sanitize@url \@href}%
\providecommand \@href[1]{\@@startlink{#1}\@@href}%
\providecommand \@@href[1]{\endgroup#1\@@endlink}%
\providecommand \@sanitize@url [0]{\catcode `\\12\catcode `\$12\catcode
  `\&12\catcode `\#12\catcode `\^12\catcode `\_12\catcode `\%12\relax}%
\providecommand \@@startlink[1]{}%
\providecommand \@@endlink[0]{}%
\providecommand \url  [0]{\begingroup\@sanitize@url \@url }%
\providecommand \@url [1]{\endgroup\@href {#1}{\urlprefix }}%
\providecommand \urlprefix  [0]{URL }%
\providecommand \Eprint [0]{\href }%
\providecommand \doibase [0]{http://dx.doi.org/}%
\providecommand \selectlanguage [0]{\@gobble}%
\providecommand \bibinfo  [0]{\@secondoftwo}%
\providecommand \bibfield  [0]{\@secondoftwo}%
\providecommand \translation [1]{[#1]}%
\providecommand \BibitemOpen [0]{}%
\providecommand \bibitemStop [0]{}%
\providecommand \bibitemNoStop [0]{.\EOS\space}%
\providecommand \EOS [0]{\spacefactor3000\relax}%
\providecommand \BibitemShut  [1]{\csname bibitem#1\endcsname}%
\let\auto@bib@innerbib\@empty
\bibitem [{\citenamefont {Grassellino}\ \emph {et~al.}(2013)\citenamefont
  {Grassellino}, \citenamefont {Romanenko}, \citenamefont {Sergatskov},
  \citenamefont {Melnychuk}, \citenamefont {Trenikhina}, \citenamefont
  {Crawford}, \citenamefont {Rowe}, \citenamefont {Wong}, \citenamefont
  {Khabiboulline},\ and\ \citenamefont {Barkov}}]{Grassellino2013b}%
  \BibitemOpen
  \bibfield  {author} {\bibinfo {author} {\bibfnamefont {A.}~\bibnamefont
  {Grassellino}}, \bibinfo {author} {\bibfnamefont {A.}~\bibnamefont
  {Romanenko}}, \bibinfo {author} {\bibfnamefont {D.}~\bibnamefont
  {Sergatskov}}, \bibinfo {author} {\bibfnamefont {O.}~\bibnamefont
  {Melnychuk}}, \bibinfo {author} {\bibfnamefont {Y.}~\bibnamefont
  {Trenikhina}}, \bibinfo {author} {\bibfnamefont {A.}~\bibnamefont
  {Crawford}}, \bibinfo {author} {\bibfnamefont {A.}~\bibnamefont {Rowe}},
  \bibinfo {author} {\bibfnamefont {M.}~\bibnamefont {Wong}}, \bibinfo {author}
  {\bibfnamefont {T.}~\bibnamefont {Khabiboulline}}, \ and\ \bibinfo {author}
  {\bibfnamefont {F.}~\bibnamefont {Barkov}},\ }\href {\doibase
  10.1088/0953-2048/26/10/102001} {\bibfield  {journal} {\bibinfo  {journal}
  {Supercond. Sci. Technol.}\ }\textbf {\bibinfo {volume} {26}},\ \bibinfo
  {pages} {102001} (\bibinfo {year} {2013})}\BibitemShut {NoStop}%
\bibitem [{\citenamefont {{Edwards (ed.)}}(1995)}]{Edwards1995}%
  \BibitemOpen
  \bibfield  {author} {\bibinfo {author} {\bibfnamefont {D.~A.}\ \bibnamefont
  {{Edwards (ed.)}}},\ }\href@noop {} {\emph {\bibinfo {title} {Concept. Des.
  Rep.}}},\ \bibinfo {type} {Tech. Rep.}\ \bibinfo {number} {Ch. 4}\ (\bibinfo
  {year} {1995})\BibitemShut {NoStop}%
\bibitem [{\citenamefont {Martinello}\ \emph {et~al.}(2015)\citenamefont
  {Martinello}, \citenamefont {Checchin}, \citenamefont {Grassellino},
  \citenamefont {Melnychuk}, \citenamefont {Posen}, \citenamefont {Romanenko},
  \citenamefont {Sergatskov},\ and\ \citenamefont
  {Zasadzinski}}]{Martinello2015}%
  \BibitemOpen
  \bibfield  {author} {\bibinfo {author} {\bibfnamefont {M.}~\bibnamefont
  {Martinello}}, \bibinfo {author} {\bibfnamefont {M.}~\bibnamefont
  {Checchin}}, \bibinfo {author} {\bibfnamefont {A.}~\bibnamefont
  {Grassellino}}, \bibinfo {author} {\bibfnamefont {O.}~\bibnamefont
  {Melnychuk}}, \bibinfo {author} {\bibfnamefont {S.}~\bibnamefont {Posen}},
  \bibinfo {author} {\bibfnamefont {A.}~\bibnamefont {Romanenko}}, \bibinfo
  {author} {\bibfnamefont {D.}~\bibnamefont {Sergatskov}}, \ and\ \bibinfo
  {author} {\bibfnamefont {J.~F.}\ \bibnamefont {Zasadzinski}},\ }\href@noop {}
  {\bibfield  {journal} {\bibinfo  {journal} {Proc. Seventeenth Int. Conf. RF
  Supercond.}\ }\textbf {\bibinfo {volume} {MOPB015}} (\bibinfo {year}
  {2015})}\BibitemShut {NoStop}%
\bibitem [{\citenamefont {Gonnella}\ \emph {et~al.}(2015)\citenamefont
  {Gonnella}, \citenamefont {Kaufman},\ and\ \citenamefont
  {Liepe}}]{Gonnella2015}%
  \BibitemOpen
  \bibfield  {author} {\bibinfo {author} {\bibfnamefont {D.}~\bibnamefont
  {Gonnella}}, \bibinfo {author} {\bibfnamefont {J.}~\bibnamefont {Kaufman}}, \
  and\ \bibinfo {author} {\bibfnamefont {M.}~\bibnamefont {Liepe}},\ }\href
  {http://arxiv.org/abs/1509.04127} {\  (\bibinfo {year} {2015})},\ \Eprint
  {http://arxiv.org/abs/1509.04127} {arXiv:1509.04127} \BibitemShut {NoStop}%
\bibitem [{\citenamefont {Padamsee}\ \emph {et~al.}(2008)\citenamefont
  {Padamsee}, \citenamefont {Knobloch},\ and\ \citenamefont
  {Hays}}]{Padamsee2008}%
  \BibitemOpen
  \bibfield  {author} {\bibinfo {author} {\bibfnamefont {H.}~\bibnamefont
  {Padamsee}}, \bibinfo {author} {\bibfnamefont {J.}~\bibnamefont {Knobloch}},
  \ and\ \bibinfo {author} {\bibfnamefont {T.}~\bibnamefont {Hays}},\ }\href
  {http://books.google.com/books/about/RF\_Superconductivity\_for\_Accelerators.html?id=OO1AAQAAIAAJ\&pgis=1}
  {\emph {\bibinfo {title} {{RF superconductivity for accelerators}}}}\
  (\bibinfo  {publisher} {Wiley-VCH},\ \bibinfo {address} {New York},\ \bibinfo
  {year} {2008})\ p.\ \bibinfo {pages} {521}\BibitemShut {NoStop}%
\bibitem [{\citenamefont {Vallet}\ \emph {et~al.}(1992)\citenamefont {Vallet},
  \citenamefont {Bolore}, \citenamefont {Bonin}, \citenamefont {Charrier},
  \citenamefont {Daillant}, \citenamefont {Gratadour}, \citenamefont
  {Koechlin},\ and\ \citenamefont {Safa}}]{Vallet1992}%
  \BibitemOpen
  \bibfield  {author} {\bibinfo {author} {\bibfnamefont {C.}~\bibnamefont
  {Vallet}}, \bibinfo {author} {\bibfnamefont {M.}~\bibnamefont {Bolore}},
  \bibinfo {author} {\bibfnamefont {B.}~\bibnamefont {Bonin}}, \bibinfo
  {author} {\bibfnamefont {J.~P.}\ \bibnamefont {Charrier}}, \bibinfo {author}
  {\bibfnamefont {B.}~\bibnamefont {Daillant}}, \bibinfo {author}
  {\bibfnamefont {J.}~\bibnamefont {Gratadour}}, \bibinfo {author}
  {\bibfnamefont {F.}~\bibnamefont {Koechlin}}, \ and\ \bibinfo {author}
  {\bibfnamefont {H.}~\bibnamefont {Safa}},\ }\href@noop {} {\bibfield
  {journal} {\bibinfo  {journal} {Proc. EPAC 1992}\ ,\ \bibinfo {pages} {1295}}
  (\bibinfo {year} {1992})}\BibitemShut {NoStop}%
\bibitem [{\citenamefont {Romanenko}\ \emph
  {et~al.}(2014{\natexlab{a}})\citenamefont {Romanenko}, \citenamefont
  {Grassellino}, \citenamefont {Melnychuk},\ and\ \citenamefont
  {Sergatskov}}]{Romanenko2014}%
  \BibitemOpen
  \bibfield  {author} {\bibinfo {author} {\bibfnamefont {A.}~\bibnamefont
  {Romanenko}}, \bibinfo {author} {\bibfnamefont {A.}~\bibnamefont
  {Grassellino}}, \bibinfo {author} {\bibfnamefont {O.}~\bibnamefont
  {Melnychuk}}, \ and\ \bibinfo {author} {\bibfnamefont {D.~A.}\ \bibnamefont
  {Sergatskov}},\ }\href {\doibase 10.1063/1.4875655} {\bibfield  {journal}
  {\bibinfo  {journal} {J. Appl. Phys.}\ }\textbf {\bibinfo {volume} {115}},\
  \bibinfo {pages} {184903} (\bibinfo {year} {2014}{\natexlab{a}})}\BibitemShut
  {NoStop}%
\bibitem [{\citenamefont {Romanenko}\ \emph
  {et~al.}(2014{\natexlab{b}})\citenamefont {Romanenko}, \citenamefont
  {Grassellino}, \citenamefont {Crawford}, \citenamefont {Sergatskov},\ and\
  \citenamefont {Melnychuk}}]{Romanenko2014a}%
  \BibitemOpen
  \bibfield  {author} {\bibinfo {author} {\bibfnamefont {A.}~\bibnamefont
  {Romanenko}}, \bibinfo {author} {\bibfnamefont {A.}~\bibnamefont
  {Grassellino}}, \bibinfo {author} {\bibfnamefont {A.~C.}\ \bibnamefont
  {Crawford}}, \bibinfo {author} {\bibfnamefont {D.~A.}\ \bibnamefont
  {Sergatskov}}, \ and\ \bibinfo {author} {\bibfnamefont {O.}~\bibnamefont
  {Melnychuk}},\ }\href {\doibase 10.1063/1.4903808} {\bibfield  {journal}
  {\bibinfo  {journal} {Appl. Phys. Lett.}\ }\textbf {\bibinfo {volume}
  {105}},\ \bibinfo {pages} {234103} (\bibinfo {year}
  {2014}{\natexlab{b}})}\BibitemShut {NoStop}%
\bibitem [{\citenamefont {Huebener}\ and\ \citenamefont
  {Seher}(1969)}]{Huebener1969}%
  \BibitemOpen
  \bibfield  {author} {\bibinfo {author} {\bibfnamefont {R.~P.}\ \bibnamefont
  {Huebener}}\ and\ \bibinfo {author} {\bibfnamefont {A.}~\bibnamefont
  {Seher}},\ }\href {\doibase 10.1103/PhysRev.185.666} {\bibfield  {journal}
  {\bibinfo  {journal} {Phys. Rev.}\ }\textbf {\bibinfo {volume} {185}},\
  \bibinfo {pages} {666} (\bibinfo {year} {1969})}\BibitemShut {NoStop}%
\bibitem [{\citenamefont {Kubo}(2016)}]{Kubo2016}%
  \BibitemOpen
  \bibfield  {author} {\bibinfo {author} {\bibfnamefont {T.}~\bibnamefont
  {Kubo}},\ }\href {http://arxiv.org/abs/1601.02118} {\  (\bibinfo {year}
  {2016})},\ \Eprint {http://arxiv.org/abs/1601.02118} {arXiv:1601.02118}
  \BibitemShut {NoStop}%
\bibitem [{\citenamefont {Dasgupta}\ \emph {et~al.}(1978)\citenamefont
  {Dasgupta}, \citenamefont {Koch}, \citenamefont {Kroeger},\ and\
  \citenamefont {Chou}}]{Dasgupta1978}%
  \BibitemOpen
  \bibfield  {author} {\bibinfo {author} {\bibfnamefont {A.}~\bibnamefont
  {Dasgupta}}, \bibinfo {author} {\bibfnamefont {C.~C.}\ \bibnamefont {Koch}},
  \bibinfo {author} {\bibfnamefont {D.~M.}\ \bibnamefont {Kroeger}}, \ and\
  \bibinfo {author} {\bibfnamefont {Y.~T.}\ \bibnamefont {Chou}},\ }\href
  {\doibase 10.1080/13642817808245338} {\bibfield  {journal} {\bibinfo
  {journal} {Philos. Mag. Part B}\ }\textbf {\bibinfo {volume} {38}},\ \bibinfo
  {pages} {367} (\bibinfo {year} {1978})}\BibitemShut {NoStop}%
\bibitem [{\citenamefont {Santhanam}(1976)}]{Santhanam1976}%
  \BibitemOpen
  \bibfield  {author} {\bibinfo {author} {\bibfnamefont {A.~T.}\ \bibnamefont
  {Santhanam}},\ }\href {\doibase 10.1007/BF02396644} {\bibfield  {journal}
  {\bibinfo  {journal} {J. Mater. Sci.}\ }\textbf {\bibinfo {volume} {11}},\
  \bibinfo {pages} {1099} (\bibinfo {year} {1976})}\BibitemShut {NoStop}%
\bibitem [{Note1()}]{Note1}%
  \BibitemOpen
  \bibinfo {note} {The calculation of $B_{SC}/B_{NC}$ for full expulsion takes
  into account the thickness of the fluxgate probe but not its length, and it
  assumes that the probe is ideally centered at the equator. Calculations show
  that the effect of these factors is expected to be $<5\%$, and they are
  omitted in the figures for simplicity.}\BibitemShut {Stop}%
\bibitem [{Note2()}]{Note2}%
  \BibitemOpen
  \bibinfo {note} {Sources of measurement uncertainty in $\Delta T$ include
  thermal impedance between cavity and thermometer, assumed to result in
  overall uncertainty of 0.4 K. Sources of measurement uncertainty in
  $B_{SC}/B_{NC}$ include radial background fields and misalignment of fluxgate
  probe, assumed to result in overall uncertainty of 0.1.}\BibitemShut {Stop}%
\bibitem [{Note3()}]{Note3}%
  \BibitemOpen
  \bibinfo {note} {Higher starting temperatures generally led to larger $\Delta
  T$}\BibitemShut {NoStop}%
\bibitem [{\citenamefont {Aull}\ \emph {et~al.}(2012)\citenamefont {Aull},
  \citenamefont {Kugeler},\ and\ \citenamefont {Knobloch}}]{Aull2012}%
  \BibitemOpen
  \bibfield  {author} {\bibinfo {author} {\bibfnamefont {S.}~\bibnamefont
  {Aull}}, \bibinfo {author} {\bibfnamefont {O.}~\bibnamefont {Kugeler}}, \
  and\ \bibinfo {author} {\bibfnamefont {J.}~\bibnamefont {Knobloch}},\ }\href
  {\doibase 10.1103/PhysRevSTAB.15.062001} {\bibfield  {journal} {\bibinfo
  {journal} {Phys. Rev. Spec. Top. - Accel. Beams}\ }\textbf {\bibinfo {volume}
  {15}},\ \bibinfo {pages} {062001} (\bibinfo {year} {2012})}\BibitemShut
  {NoStop}%
\bibitem [{\citenamefont {Dhavale}\ \emph {et~al.}(2012)\citenamefont
  {Dhavale}, \citenamefont {Dhakal}, \citenamefont {Polyanskii},\ and\
  \citenamefont {Ciovati}}]{Dhavale2012}%
  \BibitemOpen
  \bibfield  {author} {\bibinfo {author} {\bibfnamefont {A.~S.}\ \bibnamefont
  {Dhavale}}, \bibinfo {author} {\bibfnamefont {P.}~\bibnamefont {Dhakal}},
  \bibinfo {author} {\bibfnamefont {A.~a.}\ \bibnamefont {Polyanskii}}, \ and\
  \bibinfo {author} {\bibfnamefont {G.}~\bibnamefont {Ciovati}},\ }\href
  {\doibase 10.1088/0953-2048/25/6/065014} {\bibfield  {journal} {\bibinfo
  {journal} {Supercond. Sci. Technol.}\ }\textbf {\bibinfo {volume} {25}},\
  \bibinfo {pages} {065014} (\bibinfo {year} {2012})}\BibitemShut {NoStop}%
\bibitem [{Note4()}]{Note4}%
  \BibitemOpen
  \bibinfo {note} {The substantial $Q$-slope observed after furnace treatment
  is not expected to be related to the flux expulsion mechanism (based on
  similar cavities that expelled well) nor to the grain growth (based on
  experience with large grain cavities). Rather, it is believed to be due to
  contamination from the furnace, based on observations of other cavities
  treated in a similar timeframe.}\BibitemShut {Stop}%
\end{thebibliography}%

\end{document}